\documentclass[conference]{IEEEtran}
\IEEEoverridecommandlockouts
\usepackage{cite}
\usepackage{amsmath,amssymb,amsfonts}
\usepackage{booktabs}
\usepackage{array,tabularx,multirow}
\usepackage{graphicx}
\usepackage[caption=false,font=footnotesize]{subfig}
\usepackage{textcomp}
\usepackage{xcolor}
\usepackage{url}
\usepackage{enumitem}
\usepackage{float}
\usepackage{placeins}
\usepackage{balance}
\usepackage{siunitx}
\usepackage{tikz}
\usetikzlibrary{arrows.meta,positioning,fit,backgrounds,shapes,calc}
\usepackage{amsthm}
\usepackage[linesnumbered,ruled,vlined]{algorithm2e}
\usepackage[hidelinks]{hyperref}

\definecolor{revisionblue}{RGB}{0,0,0}
\newcommand{\rev}[1]{#1}
\newcommand{\mscore}{\mathcal{M}}
\newcommand{\amscore}{\widetilde{\mathcal{M}}}
\newcommand{\stitle}[1]{\par\addvspace{1ex}\noindent\textbf{#1}}
\newcommand{\nop}[1]{}

\newcolumntype{L}[1]{>{\raggedright\arraybackslash}p{#1}}
\newcolumntype{C}[1]{>{\centering\arraybackslash}p{#1}}
\newcolumntype{R}[1]{>{\raggedleft\arraybackslash}p{#1}}

\SetCommentSty{mycommfont}
\SetKwRepeat{Do}{do}{while}
\newtheorem{lemma}{Lemma}[section]

\newtheorem{proposition}{Proposition}[section]

\theoremstyle{remark}
\newtheorem{remark}{Remark}[section]
\theoremstyle{definition}
\newtheorem{definition}{Definition}

\graphicspath{{figures/}}

\title{SIMQ: A Single Graph-based Index for Multi-Vector Query}

\author{
Yifan Lu{\small$^{1}$}, Jiabao Jin{\small$^{1}$}, Xiaoyao Zhong{\small$^{1}$},  Peng Cheng{\small$^{1}$}, Xiangyu Wang{\small$^{2}$}, Haotian Li{\small$^{2}$}, Hua Chen{\small$^{2}$}, \\
Junqing Chen{\small$^{2}$}, Wei Xi{\small$^{3}$}, Heng Tao Shen{\small$^{1}$}\\
\fontsize{10}{10}\itshape
$^{1}$Tongji University, Shanghai, China; $^{2}$Ant Group, Shanghai, China; $^{3}$Xi'an Jiaotong University, Xi'an, China\\
\fontsize{9}{9}\upshape
2353549@tongji.edu.cn, jiabaojin0723@gmail.com, xiaoyao.zhong@hotmail.com, cspcheng@tongji.edu.cn,\\
\{wxy407827, tianlan.lht, xuanji.ch, junqing.cjq\}@antgroup.com, xiwei@xjtu.edu.cn, shenhengtao@hotmail.com
}
\begin{document}
\maketitle
\begin{abstract}
Multi-vector retrieval models such as ColBERT achieve state-of-the-art
effectiveness by representing documents as sets of token-level vectors and scoring
via MaxSim, but their high computational cost limits scalability to large corpora.
Existing approximate methods face fundamental bottlenecks: centroid interaction
matrix approaches require computing a full matrix at every query and rely on
GPU-accelerated clustering, while data-oblivious compression methods yield
unstable recall that requires repeated compensation. Graph-based approaches may
retain existing centroid-clustering and residual-compression pipelines, operate at
the document-set level with complex construction procedures, or demand specialized
GPU hardware. We propose \textbf{SIMQ} (\textbf{S}ingle Graph
\textbf{I}ndex for \textbf{M}ulti-vector \textbf{Q}uery), a framework that bridges
single-vector graph indices and multi-vector retrieval through data-aware
sampling-based clustering and per-token graph navigation, simultaneously bypassing
all aforementioned bottlenecks. SIMQ organizes all token vectors into fine-grained
clusters via data-aware sampling, builds a single graph index over cluster
centroids independently of PLAID's residual representation, and approximates the MaxSim score through per-token graph search without
materializing any interaction matrix or relying on GPU acceleration. We provide
theoretical analysis showing that, under a manifold assumption and successful cell
retrieval, data-aware partitioning has a smaller expected
approximation error than a data-oblivious partition. Building on this foundation,
SIMQ+ further introduces count-based
scoring and a dynamic candidate set to improve robustness across varying query
conditions. Experiments on five benchmarks compare recall, throughput, and construction
time against several baselines, primarily on CPU; large-scale PLAID $k$-means centroids
are built on a separate GPU machine.
SIMQ has been integrated into VSAG,
an Ant Group's open-source vector search library.
\end{abstract}

\begin{IEEEkeywords}
multi-vector retrieval, approximate nearest neighbor search, data-aware partition
\end{IEEEkeywords}

\section{Introduction}
\label{sec:introduction}

\begin{figure}[t]
	\centering
	\includegraphics[width=\linewidth]{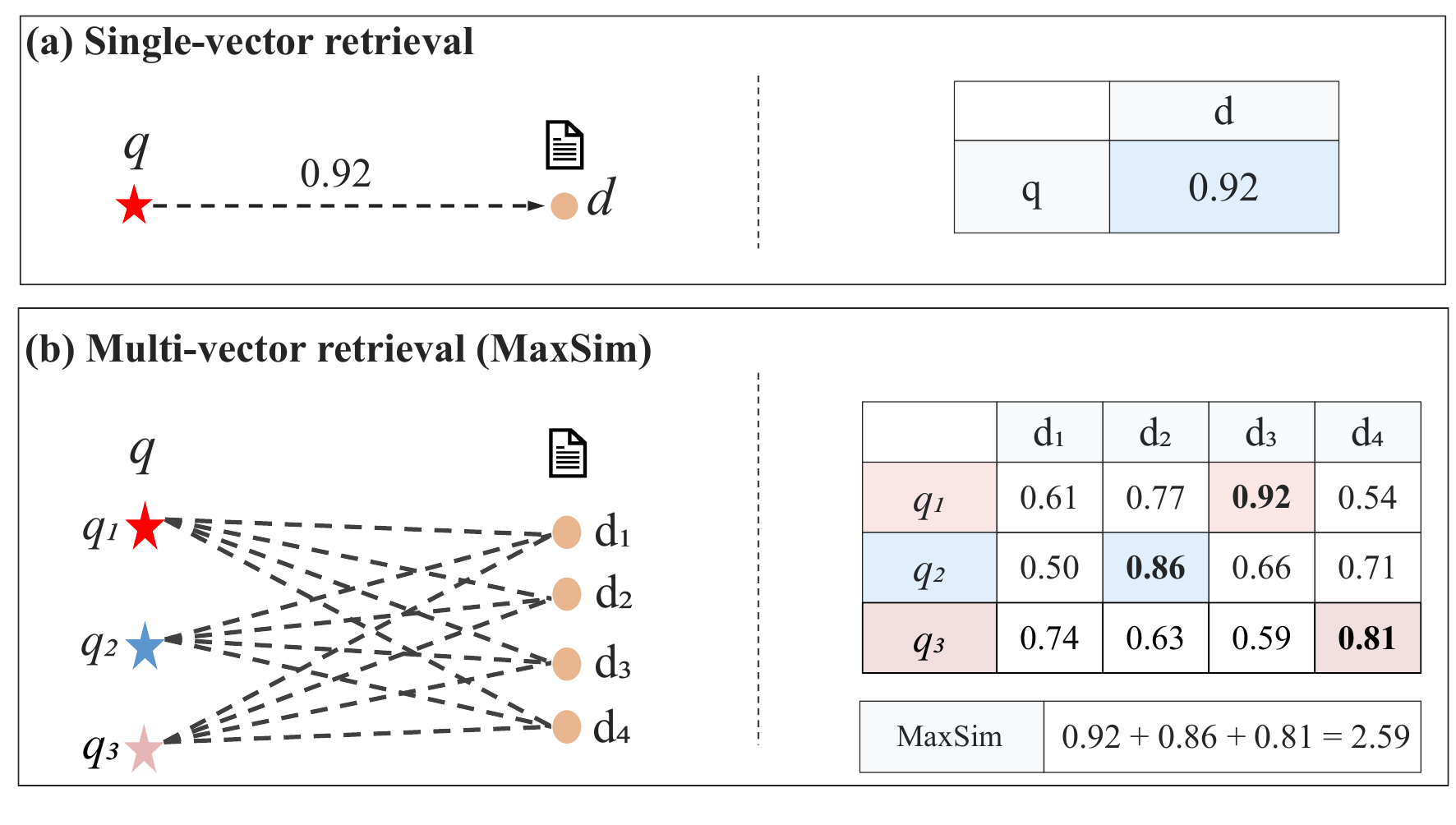}\vspace{-2ex}
	\caption{Multi-vector retrieval represents each document as a set of token-level
		vectors and scores a query through late-interaction MaxSim.}\vspace{-4ex}
	\label{fig:multivector}
\end{figure}

Multi-vector retrieval has emerged as a highly effective paradigm for dense passage
retrieval, achieving state-of-the-art performance across a wide range of
benchmarks~\cite{khattab2020colbert,santhanam2022colbertv2}. As illustrated in Fig.~\ref{fig:multivector}, unlike single-vector
models that compress an entire document into one embedding, multi-vector models, such
as ColBERT~\cite{khattab2020colbert}, produce contextualized token-level embeddings, capturing richer semantic structure. Then, to estimate the similarity between two multi-vectors, the MaxSim scoring mechanism is commonly used, however, whose exact evaluation  computation complexity is  prohibitive at scale. 

Existing multi-vector retrieval methods each face fundamental limitations rooted in their partitioning
strategy. PLAID's~\cite{santhanam2022plaid} centroid-based approach, although data-aware in its $k$-means
clustering, is forced by the interaction-matrix bottleneck to keep the number of
centroids $K$ small; each centroid must then cover a large, semantically diverse
region that may span multiple clusters, and running $k$-means over hundreds of
millions of token vectors further requires GPU acceleration.
IGP~\cite{bian2025igp} adds graph navigation over PLAID's existing clustering
centroids, but retains the underlying clustering and residual-compressed token
representation.
MUVERA's~\cite{dhulipala2024muvera} LSH
partitioning, by contrast, is data-oblivious: the random hyperplanes slice the space, which mixes semantically distinct tokens into the
same cell. Repeating the pipeline many times
compensates for the resulting recall variability.


Designing an efficient multi-vector retrieval framework
faces three key challenges:
\textit{(C1) Data-aware partitioning without GPU.}
Producing fine-grained, semantically coherent clusters typically requires $k$-means
clustering over hundreds of millions of token vectors, which is prohibitively slow
on CPU and depends on GPU acceleration. A partitioning strategy must remain data-aware
while avoiding iterative convergence.
\textit{(C2) Avoiding the interaction matrix bottleneck.}
Any approach that computes a full centroid-query interaction matrix at every query
faces a fundamental trade-off: reducing the number of centroids hurts cluster
granularity, while increasing it raises the per-query cost proportionally. The
framework must bypass the interaction matrix entirely while retaining fine-grained
clustering.
\textit{(C3) Robust approximate scoring.}
Under a fine-grained partition, centroid inner products are uniformly close to 1
for all retrieved centroids, making the approximate scores of different candidate
documents similar in magnitude. A robust scoring mechanism must separate relevant
documents from irrelevant ones under such near-uniform centroid similarity.

In this paper, we propose SIMQ (\underline{s}ingle graph \underline{i}ndex for
\underline{m}ulti-vector \underline{q}uery), a framework that bridges single-vector graph
indices and multi-vector retrieval through a data-aware centroid graph. SIMQ organizes all token vectors into fine-grained
clusters via sampling-based data-aware partitioning, builds a single graph index over
the centroids, and computes the approximate score $\amscore(d,Q)$ through per-token
graph search, where $Q$ indicates a query and $d$ denotes a document. SIMQ constructs
its  token-level partition through sampling, assignment, and local splitting,
then indexes the updated cluster centroids.
SIMQ addresses (C1) with sampling and graph-guided assignment, followed by local splits
for oversized clusters, rather than corpus-wide iterative $k$-means. It addresses (C2)
by bypassing the interaction matrix. SIMQ+ addresses (C3) with count-based scoring and
a dynamic candidate set. We provide a theoretical
analysis that under the stated manifold and
successful-cell-retrieval assumptions, SIMQ's data-aware partitioning has smaller
expected approximation error  than MUVERA's
data-oblivious LSH partitioning.

To summarize, we make the following contributions:
\begin{itemize}[leftmargin=*]
    \item We identify the dual bottleneck of centroid interaction matrix methods and
    the partition instability of data-oblivious compression methods for multi-vector
    retrieval, and formalize the problem in \S\ref{sec:problemDefinition}.

    \item We propose SIMQ, a framework that applies data-aware sampling-based
    clustering and per-token graph navigation to approximate MaxSim without interaction
    matrices or GPU-accelerated clustering in \S\ref{sec:solution1}.  We provide a theoretical analysis showing that, under explicit manifold and
    successful-cell-retrieval assumptions, SIMQ's data-aware partitioning has lower
    expected intra-cell approximation error than MUVERA's data-oblivious LSH
    partitioning in \S\ref{sec:solution2}.

    \item We propose SIMQ+ with count-based scoring and a dynamic candidate set to
    further improve robustness in \S\ref{sec:solution3}.

    \item We conduct extensive experiments on five standard benchmarks to validate the
    efficiency and effectiveness of SIMQ and SIMQ+ in \S\ref{sec:experimental}.
\end{itemize}


\section{Preliminaries}
\label{sec:problemDefinition}

\subsection{Problem Definition}

Let $\mathcal{D}$ be a corpus of $N$ documents. A query $Q$ contains $m$ token
vectors, and $V(d)$ is the set of token vectors of document $d$, with
$p_d=|V(d)|$ and corpus average token size  $p=N^{-1}\sum_{d\in\mathcal{D}}p_d$.
Every token vector has dimension $h$. We use $k$ for the number of documents
returned to the user.

We note $\delta(v, v') = \|v - v'\|_2$ for the $\ell_2$ distance between vectors $v$ and $v'$, and
$\langle v,v'\rangle$ denotes their inner product. 
For a query-token vector $q$ and a document-token vector $v$, both $\ell_2$-normalized
as in ColBERT-style models, inner product and $\ell_2$ distance satisfy:
{\small\begin{equation}
  \langle q, v \rangle = 1 - \tfrac{1}{2}\,\delta^2(q, v).
  \label{eq:ip_dist}
\end{equation}}

\begin{definition}[Multi-Vector Retrieval]
\label{def:mvretrieval}
Given a query $Q = \{q_1, \ldots, q_m\} \subset \mathbb{R}^h$ and a document corpus
$\mathcal{D} = \{d_1, \ldots, d_N\}$, where each document $d$ is represented by a set
of  vectors $V(d) = \{v_1, \ldots, v_{p_d}\} \subset \mathbb{R}^h$, the MaxSim
score between $Q$ and $d$ measures their semantic similarity. In \emph{inner product form}, it is defined as:
{\small\begin{equation}
  \mscore(d, Q) = \sum_{i=1}^{m} \max_{v \in V(d)} \langle q_i, v \rangle.
  \label{eq:maxsim}
\end{equation}}
\rev{Equivalently, by Eq.~\ref{eq:ip_dist}, MaxSim's \emph{distance form} is:}
{\small\begin{equation}
  \mscore(d, Q) = m - \frac{1}{2}\sum_{i=1}^{m} \min_{v \in V(d)} \delta^2(q_i, v).
  \label{eq:maxsim_dist}
\end{equation}}
The two forms (Eqs.~\ref{eq:maxsim} and~\ref{eq:maxsim_dist}) are mathematically equivalent. Throughout this paper, we use whichever form is more convenient for the context.
The goal of multi-vector retrieval is to efficiently retrieve the top-$k$ documents
with the highest MaxSim scores from $\mathcal{D}$.
\end{definition}

For a query with $m$ token vectors and a corpus of $N$ documents averaging $p$
token vectors per document, the MaxSim scoring mechanism incurs $O(m \cdot N \cdot p)$  computation complexity.

\begin{figure*}[t!]
	\centering
	\includegraphics[width=0.9\linewidth]{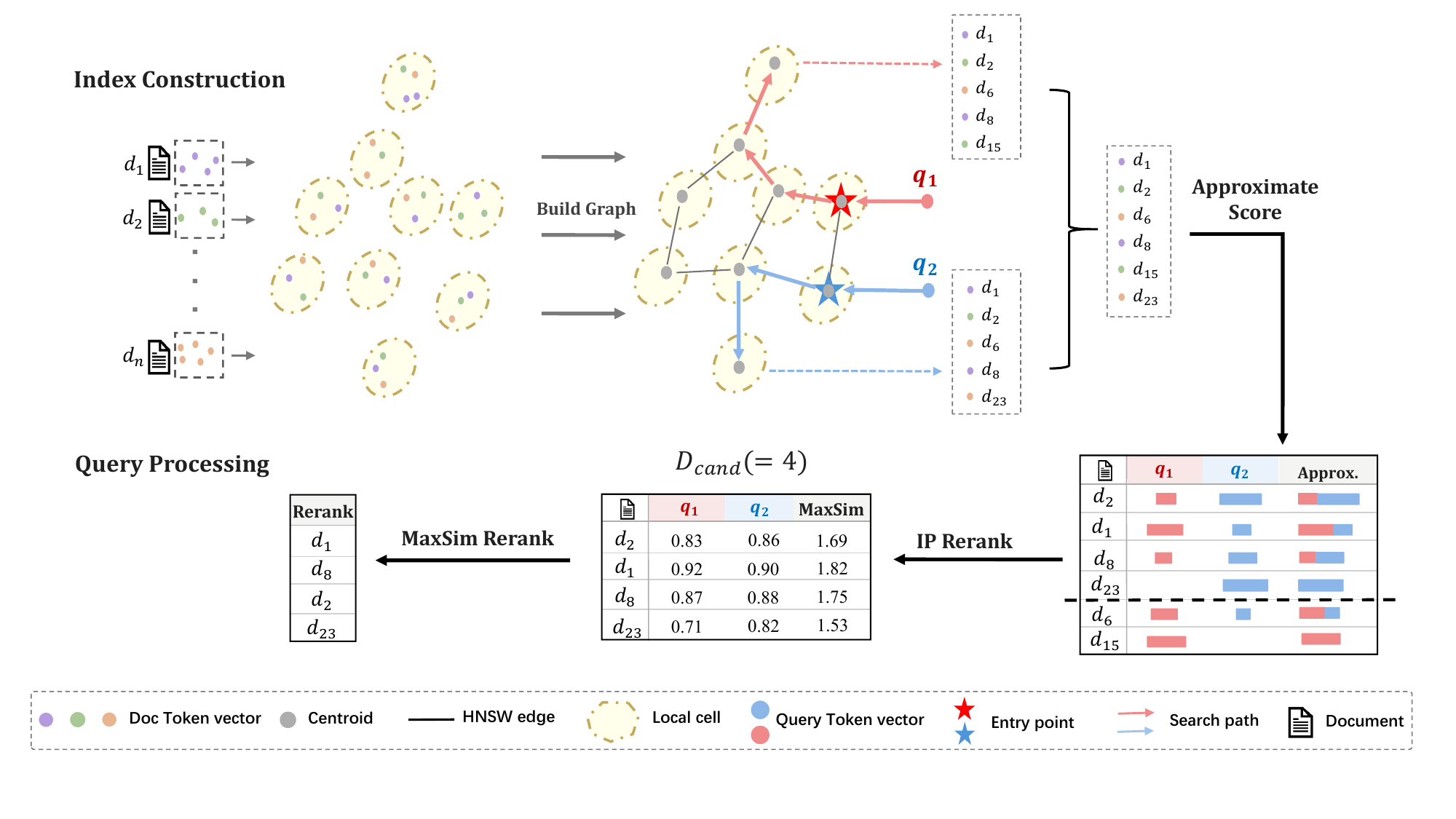}\vspace{-6ex}
	\caption{\rev{Overview of the SIMQ framework. Given a document corpus, SIMQ
			(1)~samples initial centroids, assigns token vectors, splits oversized clusters,
			and updates their centroids, (2)~builds a single query graph index over the
			updated centroids, and
			(3)~at query time, performs per-token graph navigation to collect candidate
			documents and compute approximate MaxSim scores $\amscore(d,Q)$.}}\vspace{-3ex}
	\label{fig:simq}
\end{figure*}

\subsection{Existing Multi-Vector Retrieval Approaches}

Existing methods partition token vectors, but differ in whether they preserve token-level
interactions or compress each document into a single vector, and whether their partitions adapt
to the data or not.

\noindent\textit{PLAID.} PLAID~\cite{santhanam2022plaid} applies data-aware $k$-means to
cluster the $Np$ token vectors into $K$ centroids. It preserves the multi-vector structure by
computing a $K \times m$ centroid--query-token interaction matrix, then progressively refines
the candidate documents and exactly reranks them. Residual quantization compresses token storage.
The full interaction matrix is a query-time bottleneck, while clustering very large token sets
typically requires GPU acceleration.

\noindent\textit{MUVERA.} Let $b$ be the number of binary hash bits per partition,
$d_{\mathrm{proj}}$ the projected dimension per cell, and $R$ the number of
independent partition repetitions. MUVERA~\cite{dhulipala2024muvera} uses data-oblivious LSH to
partition tokens into $2^b$ cells, then aggregates and projects each cell into a fixed-dimensional
encoding (FDE) of size $2^b d_{\mathrm{proj}}$ per document. These encodings can be indexed as
single vectors. To reduce the instability of randomized partitions, MUVERA repeats this process
$R$ times and averages the scores. The cells do not adapt to token distributions, and compressing
low-dimensional token embeddings may lose useful information. Before optional product
quantization, the FDE space is $O(R2^b d_{\mathrm{proj}}N)$, growing exponentially with $b$ and
linearly with $R$.

\section{SIMQ Framework}
\label{sec:solution1}

\subsection{Overview}

We propose SIMQ framework to efficiently solve the multi-vector retrieval problem.
Figure~\ref{fig:simq} illustrates the overall framework pipeline, which has three phases: (1)~\emph{data-aware partitioning}: all token vectors in the corpus
are organized into fine-grained clusters via a sampling-based strategy, where the
sampled vectors initialize the clusters, whose centroids are subsequently updated
after assignment and splitting; (2)~\emph{graph construction}: a single
graph index is built over the centroids, enabling efficient per-token nearest-neighbor
search at query time; (3)~\emph{query processing}: for each query token, graph search
retrieves the $n_c$ nearest centroids, and the documents associated
with these centroids form the candidate set. An approximate MaxSim score
$\amscore(d,Q)$ (defined in Eq.~\ref{eq:approx_maxsim}) is then accumulated from centroid inner products, and the top
candidates are re-ranked by exact MaxSim computation.

\subsection{Data-Aware Sampling-Based Partitioning}
\label{sec:partitioning}

SIMQ adopts a \emph{data-aware sampling-based} partitioning strategy
derived from the Point Aggregation Graph (PAG)~\cite{yu2025dsann}.
PAG is a proximity graph structure over a sampled subset of vectors, called \emph{aggregation
points}, that supports efficient nearest-neighbor navigation for the remaining \emph{residual
points}. Aggregation points are obtained via uniform random sampling and used
to construct a proximity graph; residual points are subsequently assigned to their nearest
aggregation point. PAG builds a high-quality partition
without iterative clustering, achieving significantly lower construction cost than $k$-means
while remaining entirely data-aware.

In SIMQ, we adapt PAG to the multi-vector retrieval setting: aggregation points serve as
initial centroids. After token assignment and splitting of oversized clusters, each
centroid is updated to its cluster's normalized mean. A single query graph index is
built over these updated centroids for efficient per-token navigation. 


Compared with PLAID's $k$-means clustering, the PAG-based partition has two key
advantages. First, it avoids the iterative Lloyd's algorithm that must scan all $N\cdot p$
token vectors in each iteration until convergence, reducing the build time from
$O(N\cdot p \cdot I)$ full-data iteration work (i.e., $I$ is the number of iterations)
to graph-guided assignment, with approximately logarithmic search cost per token
under fixed search settings; this comparison excludes local splitting and final graph
construction and suppresses vector dimension and centroid-comparison costs.
Second, the initial centroids are sampled directly from the data,
thus follows the empirical data distribution; by
contrast, $k$-means centroids are arithmetic means
that may fall between clusters in low-density areas, particularly when $K$ is
constrained to be small. After assignment and local splitting, SIMQ updates each
centroid to its cluster's normalized mean, then final centroids need not coincide
with observed token vectors.

\subsection{Motivation}
\label{sec:motivation}

\begin{figure}[t]
  \centering
  \includegraphics[width=\linewidth]{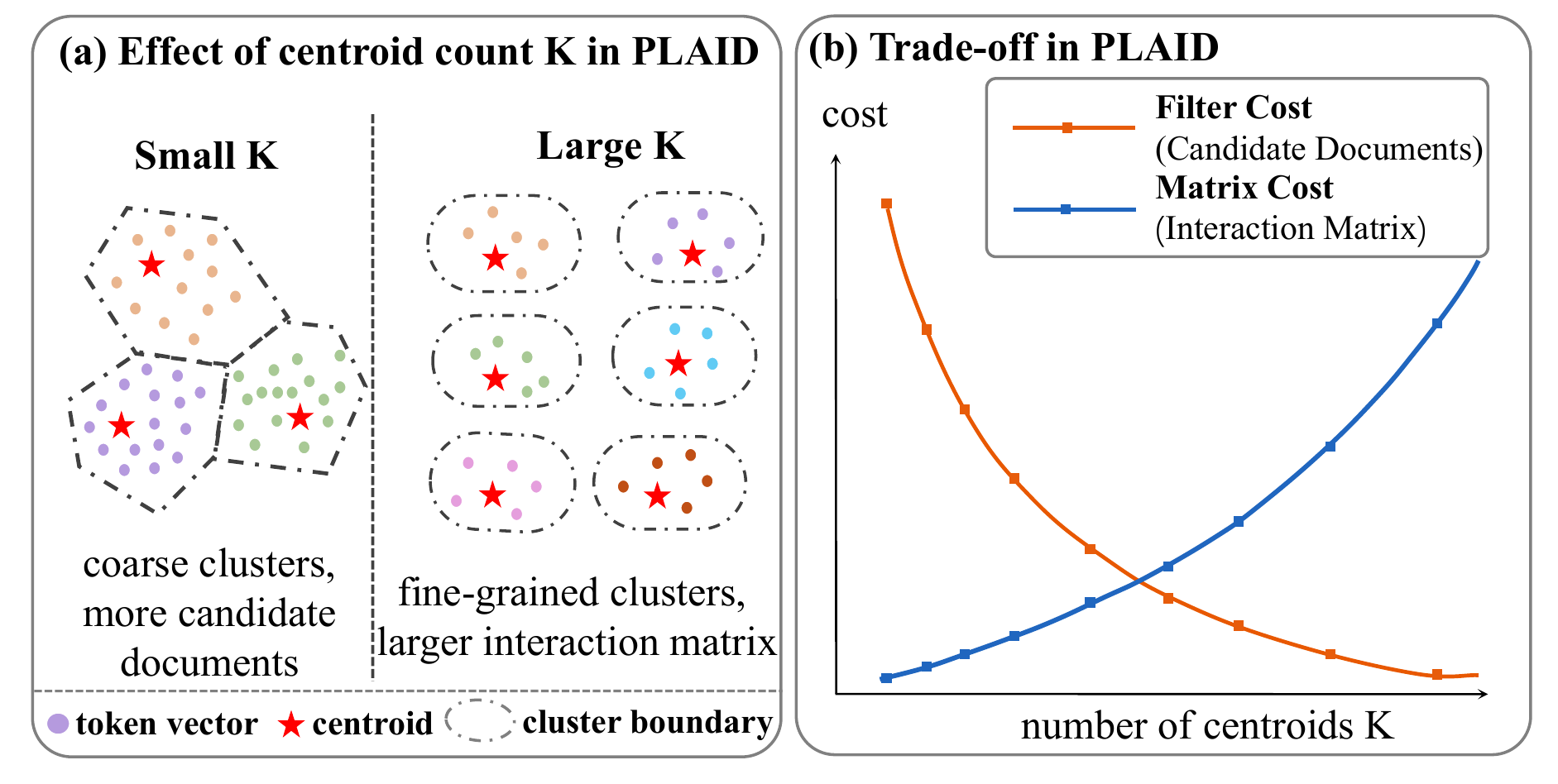}\vspace{-2ex}
  \caption{\rev{The trade-off in PLAID's number of clusters $K$.}}\vspace{-4ex}
  \label{fig:plaid_analyze}
\end{figure}

The main bottleneck of PLAID~\cite{santhanam2022plaid} is that computing the full $K \times  |Q|$ centroid–query interaction matrix at every query is mandatory. As
illustrated in Figure~\ref{fig:plaid_analyze}, changing the cluster number $K$ cannot resolve this, as decreasing $K$ increases document candidate filter costs while increasing $K$ inflates matrix computation costs. Additionally, producing PLAID's centroids via $k$-means requires processing all token vectors and typically relies on GPU acceleration at scale.


Rather than computing this full matrix, SIMQ leverages data-aware sampling and graph-based
indexing to navigate each query token directly to its most relevant centroids, naturally
bypassing irrelevant ones. This allows SIMQ to use far more centroids than PLAID without
the associated computational penalty, yielding smaller and more precise clusters while
keeping query-time cost low and avoiding GPU dependency entirely.

\subsection{Index Construction}
\label{sec:index_construction}

Given a corpus of multi-vector documents, each document $d$ is represented by a set of
single vectors $\{v_1, v_2, \ldots, v_{p_d}\} \subset \mathbb{R}^h$, where $p_d$ denotes
the number of token vectors in document $d$. We denote the average number of token vectors
per document as $p$, then the total number of single vectors across the corpus is
approximately $N\cdot p$. We construct two auxiliary mappings: $\phi(v)=d$ maps each token $v\in V(d)$
to its source document $d$, and
$\Psi(c_j)=\{\phi(v):v\in S_j\}$ maps a centroid $c_j$ to the source documents of the member-tokens
in its cluster.

\stitle{Centroid Initialization.}
Following the PAG strategy, we uniformly sample a subset of single vectors as initial
cluster centroids with sampling ratio $\alpha\in(0,1]$. Since $N\cdot p \gg \frac{1}{\alpha}$, the density
of sampled centroids converges in probability to the overall distribution of vectors in the
embedding space; in expectation, $\mathcal{C}=\alpha \cdot N\cdot p$ centroids are selected initially.

\stitle{Cluster Assignment and Splitting.}
We first build a temporary graph-index $G_{\mathrm{init}}$ over the sampled centroids.
For each remaining token vector, graph search on $G_{\mathrm{init}}$ approximately
identifies its nearest centroid for assignment. 
The online scheme described below updates the graph during assignment.
Because uniform sampling can
under-represent locally dense regions, some sampled centroids may receive many vectors from
the same region. We therefore enforce a maximum cluster size $T$. There are two ways to
enforce this bound:

\textit{Online split-insertion.} Clusters exceeding $T$ trigger immediate splitting using a farthest member as the new centroid, followed by local reassignment and incremental graph insertion. This handles local density well but couples assignment with graph mutation, requiring synchronization and limiting it to single-threaded builds. Extra insertions can also degrade construction efficiency and search quality in less robust graph backends.

\textit{Deferred recursive $k=2$ splitting.} Vectors are assigned against a fixed centroid set before oversized clusters undergo recursive binary splitting until all sizes are $\le T$. Final centroids are inserted into the graph in a single phase. Decoupling assignment from graph mutation enables safe parallelization with minimal clustering overhead.

%


\stitle{Centroid Update.}
For a cluster $j$, let $c_j$ be its centroid, $\mathcal{A}[c_j]=S_j$ be its member-token set, 
and $\mu_j$ be its unnormalized mean. Each centroid is updated as
$c_j = \mu_j / \|\mu_j\|$, where $\mu_j = \frac{1}{|S_j|} \sum_{v \in S_j} v$.
Normalization ensures all centroids lie on the unit hypersphere, consistent with the
$\ell_2$-normalized token vectors produced by ColBERT-style models.

\stitle{Graph Construction.}
SIMQ indexes the centroids through a replaceable graph-based ANNS backend.
 HNSW~\cite{malkov2020hnsw}
is our default implementation;
other graph indices (e.g.,
Vamana~\cite{subramanya2019diskann} or NSG~\cite{fu2019fast}) can be substituted
depending on the deployment scenario. 
We empirically validate SIMQ's compatibility with alternative graph structures in
\S\ref{sec:experimental}, demonstrating that the framework maintains strong
performance across different graph backends. Construction
is commonly modeled as approximately $O(|\mathcal{C}| \log |\mathcal{C}|)$
for HNSW under fixed dimension, degree, and construction-search settings, where $\mathcal{C}$ denote the set of cluster centroids.

  

\begin{algorithm}[t]\small
\caption{SIMQ Index Construction}
\label{alg:index}
\KwIn{Corpus $\mathcal{D} = \{d_1, \ldots, d_N\}$; sampling ratio $\alpha$; split threshold $T$}
\KwOut{Graph index $G$; membership map $\mathcal{A}$; mappings $\phi$, $\Psi$}
$\mathcal{V} \leftarrow$ flatten all token vectors from $\mathcal{D}$\;
$\phi(v) \leftarrow \text{doc}(v)$ for all $v \in \mathcal{V}$\;
$\mathcal{C} \leftarrow \text{UniformSample}(\mathcal{V}, \alpha)$\;
Build temporary graph $G_{\mathrm{init}}$ over $\mathcal{C}$ for token assignment\;
$\mathcal{A}[c] \leftarrow \{c\}$ for every $c\in\mathcal{C}$\;
\ForEach{$v \in \mathcal{V} \setminus \mathcal{C}$}{
    $c^* \leftarrow$ nearest centroid found by graph search on $G_{\mathrm{init}}$ for $v$\;
    $\mathcal{A}[c^*] \leftarrow \mathcal{A}[c^*] \cup \{v\}$\;
}
$\mathcal{S} \leftarrow \{\mathcal{A}[c]:c\in\mathcal{C}\}$\;
\While{some $S_j\in\mathcal{S}$ has $|S_j|>T$}{
    split $S_j$ into nonempty children $S_a,S_b$ using local $k$-means with $k=2$\;
    $\mathcal{S}\leftarrow(\mathcal{S}\setminus\{S_j\})\cup\{S_a,S_b\}$\;
}
$\mathcal{C}\leftarrow\emptyset$; \quad $\mathcal{A}\leftarrow\emptyset$\;
\ForEach{cluster $S_j\in\mathcal{S}$}{
    $c_j\leftarrow\bigl(\sum_{v\in S_j}v\bigr)/\bigl\|\sum_{v\in S_j}v\bigr\|$\;
    $\mathcal{C}\leftarrow\mathcal{C}\cup\{c_j\}$; \quad $\mathcal{A}[c_j]\leftarrow S_j$\;
}
Build graph index $G$ over $\mathcal{C}$ (e.g., using HNSW~\cite{malkov2020hnsw})\;
\ForEach{centroid $c \in \mathcal{C}$}{ $\Psi(c) \leftarrow \{\phi(v) : v \in \mathcal{A}[c]\}$ }
\KwRet{$G$, $\mathcal{A}$, $\phi$, $\Psi$}\;
\end{algorithm}

Algorithm~\ref{alg:index} details the deferred construction procedure used in our
multi-threaded implementation. In Algorithm~\ref{alg:index},
$\mathrm{doc}(v)$ returns the source document of $v$. Let
$\mathcal{V}=\bigcup_{d\in\mathcal{D}}V(d)$ denote all corpus token vectors.
The function
$\mathrm{UniformSample}(\mathcal{V},\alpha)$ samples an $\alpha$ fraction of
the token-vector collection $\mathcal{V}$. It first flattens the corpus token vectors, records
their source documents, samples initial centroids, and builds the temporary assignment
graph (Lines 1--5). The remaining vectors are assigned to their nearest retrieved
centroid, and the initial clusters are formed (Lines 6--9). Each oversized cluster is
then recursively split into two using local k-means with  $k=2$ until every cluster has at
most $T$ members (Lines 10--12). Finally, the algorithm computes normalized cluster
means, rebuilds the centroid and membership maps, constructs the final graph, and
creates the centroid-to-document map $\Psi$ (Lines 13--20). This is the deferred
strategy used in the multi-threaded experiments; online split-insertion is an
alternative described above.

\stitle{Running Example.}
Consider a small corpus of three documents $d_1, d_2, d_3$ with token vectors in a
2-dimensional embedding space. The coordinates below illustrate vector directions;
all vectors are $\ell_2$-normalized before assignment and scoring.
Suppose $d_1$ contains vectors near $(0.9, 0.1)$
(corresponding to the term ``machine''), $d_2$ contains vectors near $(0.1, 0.9)$
(corresponding to ``algorithm''), and $d_3$ contains both. With sampling ratio
$\alpha = 0.5$, two centroids $c_1 \approx (0.9, 0.1)$ and $c_2 \approx (0.1, 0.9)$
are selected. After cluster assignment, $S_1$ contains the ``machine'' vectors of
$d_1$ and $d_3$, while $S_2$ contains the ``algorithm'' vectors of $d_2$ and $d_3$.
The graph connects $c_1$ and $c_2$ (as the only two centroids), with mappings
$\Psi(c_1) = \{d_1, d_3\}$ and $\Psi(c_2) = \{d_2, d_3\}$. At query time, a query token
$q$ near $(0.85, 0.15)$ would be navigated to $c_1$ via graph search, yielding candidate
documents $\{d_1, d_3\}$ through $\Psi(c_1)$, correctly identifying the documents
containing ``machine''-related vectors.

\subsection{Query Processing}
\label{sec:query_processing}

\rev{Given a query $Q = \{q_1, \ldots, q_m\}$, SIMQ processes each query token independently to
retrieve candidate documents and compute approximate scores.

\stitle{Centroid Retrieval and Candidate Collection.}
For each query token $q_i$, we retrieve the $n_c$ nearest centroids via graph
search: $\mathcal{C}_i = \text{Search}(G, q_i, n_c)$, where $\mathrm{Search}$
returns up to $n_c$ centroid vectors from graph $G$ for query token $q_i$. The graph search starts from an entry point, then greedily traverses the graph toward the query token $q_i$ and returns the nearest centroids found. Under fixed dimension and search settings, well-constructed graphs commonly exhibit approximately logarithmic search scaling in $|\mathcal{C}|$; this is not a worst-case guarantee and does not account for increasing $n_c$. In this work, we use HNSW~\cite{malkov2020hnsw} as the underlying graph index. Candidate documents for
$q_i$ first form the raw pool $\mathcal{U}_i=\bigcup_{c\in\mathcal{C}_i}\Psi(c)$.
Let $\mathcal{C}(d)$ be the centroids whose clusters contain tokens of document $d$.
For $d\in\mathcal{U}_i$, define its per-token approximate score as
{\small\begin{equation}
  s_i(d)=\max_{c\in\mathcal{C}_i\cap\mathcal{C}(d)}\langle q_i,c\rangle.
  \label{eq:token_candidate_score}
\end{equation}}
The per-token document budget $n_{\mathrm{d}}$ limits how many documents are retained.
We note $\mathcal{D}_i=\mathrm{TopDocs}(\mathcal{U}_i,s_i,n_{\mathrm{d}})$,
where $\mathrm{TopDocs}$ keeps the highest-scoring $n_{\mathrm{d}}$ documents. 
The full candidate set is $\mathcal{D}_{\mathrm{cand}}=\bigcup_{i=1}^{m}\mathcal{D}_i$.
This design avoids
iterating through individual cluster members during query processing.

\stitle{Approximate Scoring.}
For each candidate document $d\in\mathcal{D}_{\mathrm{cand}}$, only the
per-token scores retained in $\mathcal{D}_i$ are accumulated.
 The approximate MaxSim score is
{\small\begin{equation}
  \amscore(d, Q)
    = \sum_{i:\,d\in\mathcal{D}_i} s_i(d),
  \label{eq:approx_score}
\end{equation}}
For each query token $q_i$ and document $d$, multiple retrieved clusters may
contain token vectors from $d$. SIMQ uses the centroid with the highest inner
product with $q_i$ to represent these matches. This maximum is taken over the retrieved centroids, while exact
MaxSim takes the maximum over the document's original token vectors. 
The approximation quality is analyzed formally
in \S\ref{sec:solution2}.

\begin{algorithm}[t]\small
	\caption{SIMQ Query Processing}
	\label{alg:query}
	\KwIn{Query $Q = \{q_1, \ldots, q_m\}$; Graph index $G$; mapping $\Psi$;
		access to document token vectors $V(d)$ for reranking;
		$n_c$; per-token document budget $n_{\mathrm{d}}$;
		reranking budget $k_{\mathrm{r}}$; final result count $k$}
	\KwOut{Top-$k$ documents ranked by exact MaxSim $\mscore(d, Q)$}
	\ForEach{query token $q_i \in Q$}{
		$\mathcal{C}_i \leftarrow \text{Search}(G, q_i, n_c)$\;
		$\mathcal{U}_i \leftarrow \bigcup_{c \in \mathcal{C}_i} \Psi(c)$\;
		compute $s_i(d)$ for each $d\in\mathcal{U}_i$ using Eq.~\ref{eq:token_candidate_score}\;
		$\mathcal{D}_i \leftarrow \mathrm{TopDocs}(\mathcal{U}_i,s_i,n_{\mathrm{d}})$\;
	}
	$\mathcal{D}_{\mathrm{cand}} \leftarrow \bigcup_{i=1}^{m} \mathcal{D}_i$\;
	\ForEach{$d \in \mathcal{D}_{\mathrm{cand}}$}{
		$\amscore(d, Q) \leftarrow \sum_{i:\,d\in\mathcal{D}_i} s_i(d)$\;
	}
	$\mathcal{T} \leftarrow$ top-$k_{\mathrm{r}}$ in $\mathcal{D}_{\mathrm{cand}}$ by $\amscore$\;
	\ForEach{$d \in \mathcal{T}$}{
		$\mscore(d, Q) \leftarrow \sum_{i=1}^{m} \max_{v \in V(d)} \langle q_i, v \rangle$\;
	}
	\KwRet{top-$k$ in $\mathcal{T}$ by $\mscore(d, Q)$}\;
\end{algorithm}

\stitle{Re-ranking.}
Let $\mathcal{T}$ denote the top-$k_{\mathrm{r}}$ candidates ranked by
$\amscore$, where $k_{\mathrm{r}}$ is the reranking budget and $k$ is the count of final
results. For every $d\in\mathcal{T}$, we then access its original token
vectors $V(d)$ and compute the exact MaxSim score $\mscore(d,Q)$. The system
returns the top-$k$ documents under these exact scores. This final step corrects
ranking errors that may arise when a cluster centroid is an imperfect proxy for
one or more of its member tokens.}


Algorithm~\ref{alg:query} describes the query  steps. For each query token,
SIMQ searches the centroid graph index, gathers the documents represented by the retrieved
clusters, scores each document using its best-matching retrieved centroid, and keeps
up to $n_{\mathrm{d}}$ documents (Lines 1--5). It unions the retained per-token
sets, sums their proxy scores, and selects the top-$k_{\mathrm{r}}$ candidates
(Lines 6--9). It then accesses the original token vectors of these candidates,
computes exact MaxSim, and returns the top-$k$ documents (Lines 10--12).

\section{Theoretical Analysis of SIMQ}
\label{sec:solution2}

\rev{We analyze when SIMQ's data-aware partitioning can achieve smaller MaxSim
approximation error than MUVERA's data-oblivious LSH partitioning. Throughout this section,
we use $\tilde{v}_i^{S}$ ($\varepsilon^S$) and $\tilde{v}_i^{M}$ ($\varepsilon^M$)   to denote the approximate vectors (approximation MaxSim errors)  of SIMQ
and MUVERA, respectively, and {\scriptsize$\amscore(d,Q)$} for the approximate MaxSim score.  We assume
token vectors concentrate on a low-dimensional manifold and form multiple semantic
clusters~\cite{tenenbaum2000global,bengio2013representation}. We state the additional
cluster-separation and partition-mixing assumptions needed by the comparison explicitly
below; the result is therefore conditional rather than a claim that every random LSH
partition must mix every pair of semantic clusters.}

\subsection{Approximation Framework}

\begin{definition}[Approximate Vector]
	\label{def:approx_vec}
	For a query token $q_i$ and a document $d$, let $v_i^* = \arg\max_{v \in V(d)} \langle q_i, v \rangle$ be the true best-matching vector. We define $\tilde{v}_i$ as the \emph{approximate vector} of $v_i^*$: the centroid vector  of the cluster cell containing $v_i^*$. 
\end{definition}

For example, if the query token $q_i$'s true best-matching vector $v_i^*$ is in cluster $j$, the approximate vector $\tilde{v}_i=c_j$. As Alg.~\ref{alg:query} shows, we find $\tilde{v}_i$ only from the retrieved $n_c$ centroids through graph search of query token $q_i$. Since usually $n_c$ is much smaller than the size of document tokens, the time complexity to find $\tilde{v}_i$ is also much smaller than that to calculate the true best-matching vector $v_i^*$. According to Eq.~\ref{eq:token_candidate_score}, we can denote $\tilde{v}_i=\arg\max_{c\in\mathcal{C}_i\cap\mathcal{C}(d)}\langle q_i,c\rangle$, where $\mathcal{C}_i$ is the retrieved $n_c$ centroids through graph search for query token $q_i$, and $\mathcal{C}(d)$ is the set of centroids whose corresponding clusters contains tokens of document $d$.


\begin{definition}[Approximate MaxSim Score]
	\label{def:approx_score}
	The \emph{approximate MaxSim score} is:
	{\small\begin{equation}
			\amscore(d, Q) = \sum_{i=1}^{m} \langle q_i, \tilde{v}_i \rangle
			\label{eq:approx_maxsim}
	\end{equation}}
\noindent where $m$ is the size of query tokens.
\end{definition}

\begin{definition}[Approximation MaxSim Error]
\label{def:approx_error}
The \emph{approximation MaxSim error} is:
{\small\begin{equation}
  \varepsilon = \mscore(d, Q) - \amscore(d, Q) = \sum_{i=1}^{m} \underbrace{\langle q_i, v_i^* - \tilde{v}_i \rangle}_{\varepsilon_i}
  \label{eq:approx_error}
\end{equation}}
By Cauchy--Schwarz, each per-token error satisfies:
{\small\begin{equation}
  \varepsilon_i = \langle q_i, v_i^* - \tilde{v}_i \rangle \leq \|q_i\| \cdot \delta(v_i^*, \tilde{v}_i)
  \label{eq:per_token_bound}
\end{equation}}
\end{definition}

\noindent The quality of the approximation thus depends on $\delta(v_i^*, \tilde{v}_i)$: how close the aggregation vector is to the true best-matching vector. Under the manifold and successful-cell-retrieval assumptions stated below, we compare the expected per-token error of SIMQ's data-aware partitioning with MUVERA's data-oblivious partitioning.

\subsection{Approximation Analysis}

\begin{lemma}[Reduction to Per-Token Comparison]
\label{lemma:reduction}
Suppose two methods each apply a fixed partition to all token vectors and, at query
time, successfully identify the cell containing each true best-matching vector
$v_i^*$. Then for the same query $Q$, the total approximation error comparison
reduces to per-token average error comparison:
{\small\begin{equation}
  \mathbb{E}[\varepsilon^{S}] < \mathbb{E}[\varepsilon^{M}] \iff \overline{\varepsilon^{S}} < \overline{\varepsilon^{M}}
\end{equation}}
where $\overline{\varepsilon} = \frac{1}{m}\sum_{i=1}^{m} \mathbb{E}[\varepsilon_i]$ is the per-token average error.
\end{lemma}

\begin{proof}
Since both methods apply a fixed partition to all token vectors, each $v_i^*$ is approximated by the aggregation vector of its cell under the same mechanism. By linearity of expectation:
{\small\begin{equation*}
  \mathbb{E}[\varepsilon] = \sum_{i=1}^{m} \mathbb{E}[\varepsilon_i] = m \cdot \overline{\varepsilon}
\end{equation*}}
For the same query, $m$ is identical for both methods. Therefore, $\mathbb{E}[\varepsilon^{S}] < \mathbb{E}[\varepsilon^{M}]$ if and only if $\overline{\varepsilon^{S}} < \overline{\varepsilon^{M}}$.
\end{proof}

\noindent\textbf{Interpretation.}\quad
Lemma~\ref{lemma:reduction} states that comparing the total approximation errors of two methods reduces to comparing their per-token average errors. This simplification is significant because it allows us to analyze a single query-token/document-vector pair $(q_i, v_i^*)$ in isolation, rather than reasoning about the full query-document interaction. In the remainder of this section, we focus on bounding $\mathbb{E}[\delta(v_i^*, \tilde{v}_i)]$ for a single token pair under each method.

\begin{lemma}[SIMQ Has Smaller Per-Token Error under Cluster Separation]
\label{lemma:per_token}
Under the low-dimensional-manifold assumption, suppose that SIMQ cells are contained
within semantic clusters with radius at most $r_S$, and that a MUVERA cell crosses
semantic-cluster boundaries with probability $p_{\mathrm{mix}}>0$. If the inter-cluster
separation dominates the within-cell radius, then for any query token $q_i$ and its true
best-matching vector $v_i^*$:
{\small\begin{equation}
  \mathbb{E}[\delta(v_i^*, \tilde{v}_i^{S})] < \mathbb{E}[\delta(v_i^*, \tilde{v}_i^{M})]
\end{equation}}
\end{lemma}

\begin{proof}
We analyze the two sides separately. Let $\delta_{\min}$ denote the minimum inter-cluster distance on the data manifold.

\textit{SIMQ (data-aware partitioning).} SIMQ samples centroids from the data itself and forms a Voronoi partition. By the characteristic of data-aware Voronoi partitioning~\cite{qin2019nearest}, the cell radius adapts to local data density: within a semantic cluster (a high-density region), cells are small and semantically coherent. Let $r_S$ denote the cell radius of SIMQ within a semantic cluster. Since all vectors in the cell lie within a ball of radius $r_S$, the aggregation vector $\tilde{v}_i^{S}$ (the cell centroid) satisfies:
{\small\begin{equation}
  \delta(v_i^*, \tilde{v}_i^{S}) \leq r_S
  \label{eq:simq_bound}
\end{equation}}

\textit
{MUVERA (data-oblivious partitioning).} MUVERA uses $b$ random Gaussian projections to partition the space into $2^b$ cells via SimHash. Since the partitioning hyperplanes are independent of the data distribution, they cannot adapt to semantic cluster boundaries. By the collision probability of random hyperplane LSH~\cite{charikar2002similarity}, two vectors at angle $\theta$ are separated with probability $\theta/\pi > 0$. Consequently, MUVERA cells can mix vectors from multiple semantic clusters with positive probability. Conditioned on such a mixed cell, let the cell containing $v_i^*$ intersect $s$ clusters with centers $\mu_1, \ldots, \mu_s$ satisfying $\delta(\mu_j, \mu_l) \geq \delta_{\min}$ for $j \neq l$. The aggregation vector $\tilde{v}_i^{M}$ is the mean of all vectors in the cell, which is pulled toward the centroid $\bar{\mu} = \frac{1}{s}\sum_{j=1}^{s} \mu_j$ of the mixed cluster centers. For any $v_i^*$ belonging to cluster $k$, by the triangle inequality:
{\small\begin{equation}
  \delta(v_i^*, \tilde{v}_i^{M}) \geq \delta(\mu_k, \bar{\mu}) - \delta(v_i^*, \mu_k) - \delta(\tilde{v}_i^{M}, \bar{\mu})
\end{equation}}
Since the clusters are well-separated and each has bounded internal radius, $\delta(\mu_k, \bar{\mu}) = \Omega(\delta_{\min})$ while the correction terms are $o(\delta_{\min})$. Therefore, after accounting for the mixing probability:
{\small\begin{equation}
  \mathbb{E}[\delta(v_i^*, \tilde{v}_i^{M})] = \Omega(p_{\mathrm{mix}}\delta_{\min})
  \label{eq:muvera_bound}
\end{equation}}

If $p_{\mathrm{mix}}\delta_{\min} \gg r_S$, Eqs.~\ref{eq:simq_bound} and~\ref{eq:muvera_bound} together yield:
{\small\begin{equation*}
  \mathbb{E}[\delta(v_i^*, \tilde{v}_i^{S})] \leq r_S \ll p_{\mathrm{mix}}\delta_{\min} = O\!\left(\mathbb{E}[\delta(v_i^*, \tilde{v}_i^{M})]\right)
\end{equation*}}
\end{proof}

\noindent\textbf{Interpretation.}\quad
Lemma~\ref{lemma:per_token} highlights the geometric advantage of data-aware partitioning: SIMQ's centroids ($\tilde{v}_i^S$) faithfully represent semantic clusters, whereas MUVERA's boundary crossings pull centroids ($\tilde{v}_i^M$) toward multi-cluster averages. The factor 
$p_{\mathrm{mix}}\delta_{\min}$ captures the frequency and severity of this distortion.

\begin{proposition}[Expected Approximation Error under Data-Aware Partitioning]
\label{prop:main}
Under the manifold, cluster-separation, and partition-mixing assumptions above, and assuming
that graph search identifies the cell containing each true best-matching vector, for the same
query $Q$ and document $d$:
{\small\begin{equation}
  \mathbb{E}[\varepsilon^{S}] < \mathbb{E}[\varepsilon^{M}]
  \label{eq:main_result}
\end{equation}}
\end{proposition}

\begin{proof}
Immediate from Lemma~\ref{lemma:reduction} and Lemma~\ref{lemma:per_token}: the total error comparison reduces to per-token comparison (Lemma~\ref{lemma:reduction}), and SIMQ's per-token error is strictly smaller (Lemma~\ref{lemma:per_token}).
\end{proof}

\noindent\textbf{Interpretation.}\quad
Proposition~\ref{prop:main} establishes a lower expected approximation error for SIMQ under standard assumptions. This advantage stems from SIMQ's data-aware partitioning producing semantically coherent cells, unlike MUVERA's data-oblivious LSH which mixes distinct regions. 

\begin{remark}[Limitation of MUVERA's Repeated Partitions]
MUVERA repeats its LSH partitioning $R$ times with independent hash functions and averages the resulting scores. Let $\amscore^{(r)}$ denote the approximate score under the $r$-th partition, and $\amscore^{M(R)} = \frac{1}{R}\sum_{r=1}^{R} \amscore^{(r)}$. Under independent repetitions, the law of large numbers gives:
{\small\begin{equation}
  \amscore^{M(R)} = \frac{1}{R}\sum_{r=1}^{R} \amscore^{(r)}
  \xrightarrow[R\to\infty]{} \mathbb{E}[\amscore^{(r)}]
\end{equation}}
Thus, increasing $R$ makes the result approach the average behavior of the data-oblivious partitions; it does not select or recover the best individual partition. Repetition reduces the variance caused by the random hyperplanes, but it does not remove a systematic bias caused by the mismatch between the LSH partition and the data distribution.
\end{remark}

\begin{figure}[t]
	\centering
	\includegraphics[width=\linewidth]{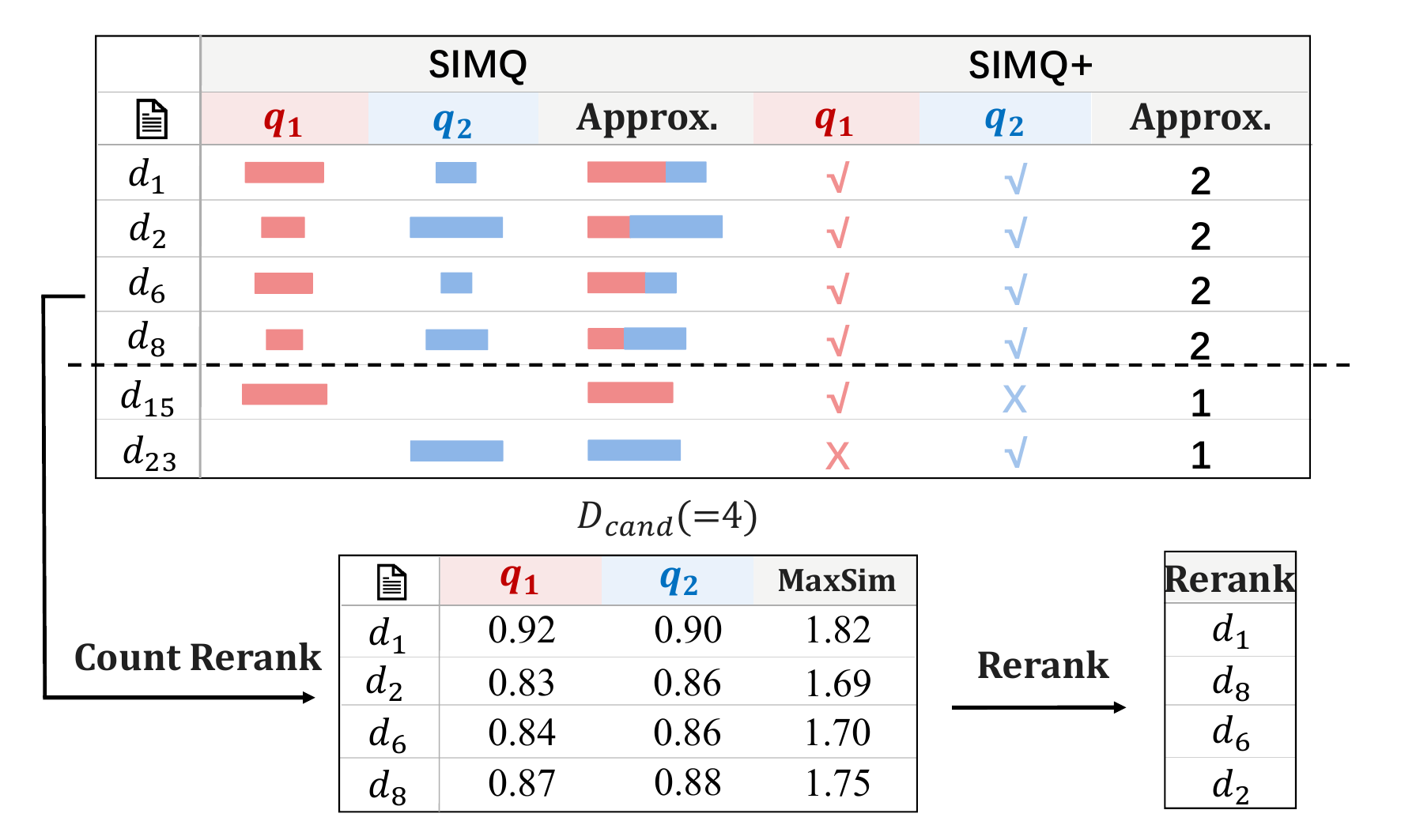}\vspace{-2ex}
	\caption{\rev{Overview of the SIMQ+ query processing.}}\vspace{-3ex}
	\label{fig:simqplus}
\end{figure}

\noindent\textbf{Interpretation.}\quad
This remark addresses a natural question: can MUVERA compensate for a weak random partition by repeating the pipeline many times? Repetition does not search for the best partition; it averages independent partitions and converges toward their expected behavior. Although this reduces variance, it cannot remove a systematic partitioning bias. This interpretation is consistent with the results in \S\ref{sec:experimental}, where MUVERA with $R = 20$ repetitions remains sensitive to the data-oblivious partitioning strategy.

\section{SIMQ+ Framework}
\label{sec:solution3}

\rev{The theoretical analysis in \S\ref{sec:solution2} shows that SIMQ's data-aware
partitioning produces fine-grained, semantically coherent cells, where each cell contains
vectors from a small, locally coherent region of a semantic cluster. Under such a
partition, the centroid of each cell closely approximates its member vectors, and query
tokens that are semantically similar to document tokens are more likely to effectively
cover them through graph search. This causes relevant documents to accumulate hits across
more query tokens, naturally separating them from irrelevant ones. SIMQ+ is designed
around this observation, introducing count-based scoring and a dynamic candidate set
to further improve retrieval quality.}

\begin{figure}[t]
  \centering
  \includegraphics[width=\linewidth]{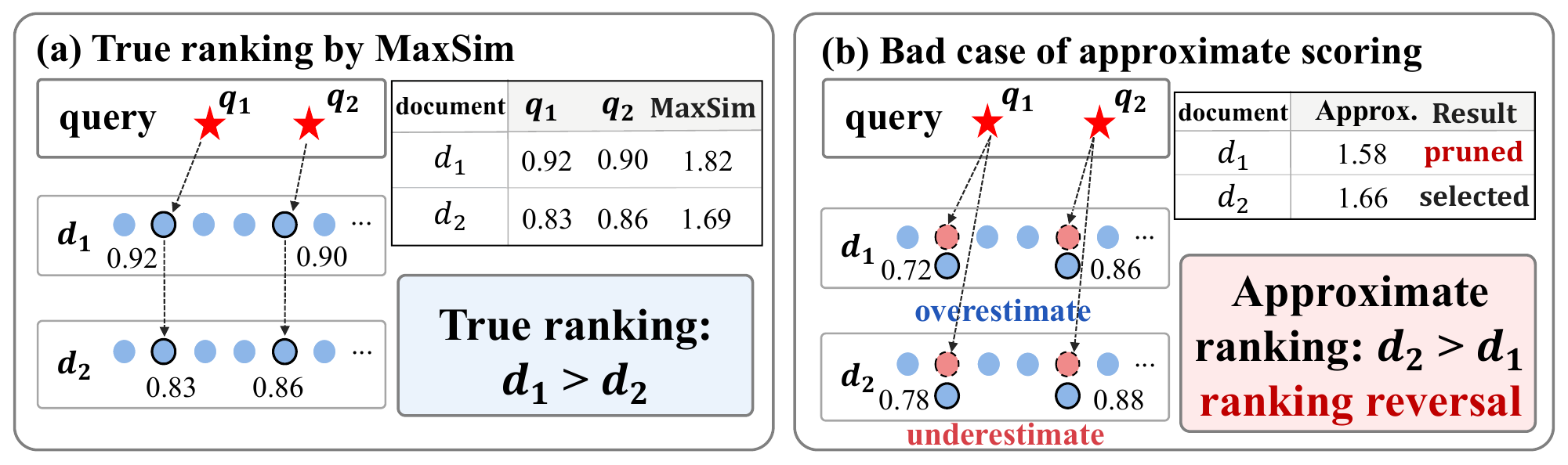}\vspace{-2ex}
  \caption{Ranking-reversal bad case motivating SIMQ+.}\vspace{-3ex}
  \label{fig:simqplus_detail}
\end{figure}

\rev{Figure~\ref{fig:simqplus} illustrates the SIMQ+ pipeline. Given a query $Q$, the
per-token centroid search and raw document-pool construction are shared with basic
SIMQ. However, SIMQ+ counts all documents in each raw pool, without per-token
approximate-document-score filtering or top-$n_{\mathrm{d}}$ truncation.
The key modifications are in the scoring and candidate selection stages: (1)~instead
of computing approximate MaxSim scores $\amscore(d,Q)$ from centroid inner products,
SIMQ+ assigns each candidate document a hit count $\text{cnt}(d)$ reflecting how many
distinct query tokens retrieved it; (2)~instead of selecting a fixed top-$k_{\mathrm{r}}$ candidate
set, SIMQ+ collects candidates in decreasing order of $\text{cnt}(d)$ and stops at
the first count level at which the accumulated candidate count reaches the minimum
reranking candidate count $N_{\min}$, retaining all ties at that level. If fewer candidates are available,
all are retained. The selected candidates are then re-ranked by exact MaxSim
computation $\mscore(d,Q)$.}

\rev{Figure~\ref{fig:simqplus_detail} illustrates the ranking-reversal bad case that motivates
SIMQ+. A relevant document with a high exact MaxSim score can be represented by an
approximate lower point, while an irrelevant document with a lower exact score can be
represented by an approximate upper point. The approximate ranking may therefore reverse
the order of the two documents before exact re-ranking. We call a
document $d$ \emph{underestimated} when its approximate score satisfies
$\amscore(d,Q) < \mscore(d,Q)$, and \emph{overestimated} when
$\amscore(d,Q) > \mscore(d,Q)$. A relevant document with a high exact MaxSim score may
therefore be pushed downward by an underestimated proxy, while an irrelevant document with
a low exact score may be promoted by an overestimated proxy. The resulting inversion is
especially harmful when a fixed candidate cutoff separates the two documents. This is the
failure mode that motivates the count-based scoring and dynamic candidate selection in SIMQ+.}

\subsection{From Inner Products to Count-Based Scoring}
\label{sec:degeneracy}

\rev{Under SIMQ's fine-grained partition, each centroid $c_j$ lies close to its member
vectors. Since all vectors are $\ell_2$-normalized, the inner product $\langle q_i, c_j \rangle$
between a query token and a nearby centroid is close to 1 for all retrieved centroids.
As a result, the approximate MaxSim scores across different candidate documents become
similar in magnitude, and the \emph{relative ordering} of documents is driven more by
\emph{how many} query tokens hit each document than by the precise inner product values.

This motivates replacing inner-product scoring with count-based scoring. For each
candidate document $d\in\bigcup_{i=1}^{m}\mathcal{U}_i$, let $\mathbf{1}[A]$ be 1 when condition $A$ holds
and 0 otherwise. We define $\text{cnt}(d) = \sum_{i=1}^{m} \mathbf{1}[d \in \mathcal{U}_i]$,
where $\mathcal{U}_i=\bigcup_{c\in\mathcal{C}_i}\Psi(c)$ contains all documents
represented in the clusters retrieved for $q_i$. Set membership ensures that a
document contributes at most one hit per query token, even if several retrieved
clusters contain its tokens. Unlike basic SIMQ, SIMQ+ does not compute $s_i(d)$
or apply $\mathrm{TopDocs}$ to this pool. Before exact reranking, centroid inner
products guide graph search but do not filter or rank the document candidates.
A document's count reflects the number of distinct query tokens for which it was covered,
providing a more robust signal than inner product values when centroids are close to their
member vectors.

Count-based scoring treats all hits equally, flattening the tail of the score
distribution and reducing the advantage of spurious high-scoring irrelevant documents
that arise from favorable centroid alignment. When the number of query tokens $m$ is
large, relevant documents accumulate hits across more tokens due to semantic relevance,
and count-based scoring naturally produces a clear separation between relevant and
irrelevant documents.

From a probabilistic perspective, the expected hit count of a document $d$ is
$\mathbb{E}[\text{cnt}(d)] = \sum_{i=1}^{m} P(d \in \mathcal{U}_i)$, where the
probability is taken over the randomness of the query. For a relevant document whose
token vectors are semantically close to many query tokens, $P(d \in \mathcal{U}_i)$
is high for a large fraction of tokens, yielding a large expected count. For an
irrelevant document, $P(d \in \mathcal{U}_i)$ is small for most tokens, yielding a
small expected count. The difference in expected counts provides a robust signal for
separating relevant from irrelevant documents, which is more reliable than the
fine-grained inner product values when centroids are close to their member vectors.}

\begin{algorithm}[t]\small
	\caption{SIMQ+ Query Processing}
	\label{alg:simqplus}
	\KwIn{Query $Q = \{q_1, \ldots, q_m\}$; Graph index $G$; mapping $\Psi$;
		access to document token vectors $V(d)$ for reranking;
		centroids retrieved per token $n_c$;
		final top-$k$; minimum candidate size $N_{\min}$}
	\KwOut{Top-$k$ documents ranked by exact MaxSim $\mscore$}
	\ForEach{query token $q_i \in Q$}{
		$\mathcal{C}_i \leftarrow \text{Search}(G,\; q_i,\; n_c)$\;
		$\mathcal{U}_i \leftarrow \bigcup_{c \in \mathcal{C}_i} \Psi(c)$\;
	}
	\ForEach{document $d$ such that $\exists\, i: d \in \mathcal{U}_i$}{
		$\text{cnt}(d) \leftarrow \sum_{i=1}^{m} \mathbf{1}[d \in \mathcal{U}_i]$\;
	}
	$\mathcal{D}_{\text{cand}} \leftarrow \bigcup_{i=1}^{m}\mathcal{U}_i$\;
	\If{$|\mathcal{D}_{\text{cand}}|\geq N_{\min}$}{
		select the largest feasible threshold $\tau$ using Eq.~\ref{eq:dynamic_candidate}\;
		$\mathcal{D}_{\text{cand}} \leftarrow \{d\in\mathcal{D}_{\text{cand}} : \text{cnt}(d) \geq \tau\}$\;
	}
	\ForEach{document $d \in \mathcal{D}_{\text{cand}}$}{
		$\mscore(d, Q) \leftarrow \sum_{i=1}^{m} \max_{v \in V(d)} \langle q_i, v \rangle$\;
	}
	\KwRet{top-$k$ documents in $\mathcal{D}_{\text{cand}}$ ranked by $\mscore(\cdot, Q)$}\;
\end{algorithm}

\subsection{Dynamic Candidate Set}
\label{sec:dynamic_candidate}

\rev{SIMQ uses a fixed top-$k_{\mathrm{r}}$ candidate set for re-ranking. SIMQ+ replaces this with a
dynamic threshold strategy: candidates are collected in decreasing order of $\text{cnt}(d)$,
and collection stops at the first count level where at least $N_{\min}$ documents
have been collected. Equivalently, $\tau$ is the largest integer threshold that
retains at least $N_{\min}$ documents. In the sets below,
$d$ ranges only over $\bigcup_{i=1}^{m}\mathcal{U}_i$, the documents hit
by at least one query token before any document truncation. When this pool contains at least $N_{\min}$ documents,
we select

{\small\begin{equation}\small
  \begin{aligned}
  \mathcal{D}_{\text{cand}} &= \{d : \text{cnt}(d) \geq \tau\},\\
  \tau &= \max_{\substack{t\in\{1,\ldots,m\}\\
    |\{d : \text{cnt}(d) \geq t\}| \geq N_{\min}}} t.
  \end{aligned}
  \label{eq:dynamic_candidate}
\end{equation}}

If the pool contains fewer than $N_{\min}$ documents, we retain the entire pool;
an empty pool yields no results. This ensures that all documents sharing the same boundary count value are included
together, preventing arbitrary exclusion of relevant documents that happen to tie at the
threshold. When enough candidates are available, the selected set has at least
$N_{\min}$ documents and may be larger when many documents share the boundary count.

The motivation for the dynamic candidate set is as follows. Under basic SIMQ, the
top-$k_{\mathrm{r}}$ candidates are selected by the approximate score $\amscore(d,Q)$. When many
documents share similar approximate scores, a fixed cutoff at $k_{\mathrm{r}}$ may split documents
with identical scores, including some while excluding others arbitrarily. By contrast,
the dynamic strategy retains all documents tied at the boundary count. When at least
$N_{\min}$ candidates are available, the selected set exceeds $N_{\min}$ only when
including the entire boundary group takes its size beyond that minimum. Thus, the
amount of expansion depends on the number of tied documents at the boundary, rather
than on high hit counts alone. This avoids arbitrarily splitting a tied group while
excluding documents with lower counts than the boundary.}


Algorithm~\ref{alg:simqplus} summarizes SIMQ+ query processing. It searches the
centroid graph for each query token and forms the corresponding document pools
(Lines 1--3). It counts how many distinct query-token pools contain each document
and forms the full candidate set (Lines 4--6). If at least $N_{\min}$ candidates
are available, it applies the dynamic threshold $\tau$, retaining all candidates
tied at the boundary (Lines 7--9). Finally, it computes exact MaxSim from the
selected documents' original token vectors and returns the top-$k$ results
(Lines 10--12).

\section{Experimental Study}
\label{sec:experimental}

In this section, we evaluate the performance of SIMQ and SIMQ+ through extensive
experiments on five benchmark datasets. We evaluate index construction and query
processing on CPU-based systems to demonstrate the practical efficiency of our approach. PLAID's large-scale
$k$-means centroids are the exception: because CPU construction was prohibitively slow,
those final centroids were generated on a separate GPU machine. We aim to answer the
following research questions:

\begin{itemize}[leftmargin=*]
    \item \textbf{RQ1:} How does SIMQ compare with state-of-the-art multi-vector retrieval methods in terms of recall, query throughput, and index construction time?

    \item \textbf{RQ2:} How does the PAG-based clustering strategy compare with alternative clustering methods (e.g., $k$-means and LSH) in terms of retrieval quality and query throughput?

    \item \textbf{RQ3:} How does SIMQ+ improve upon SIMQ, and what benefit does its count-based scoring and dynamic candidate selection provide?

    \item \textbf{RQ4:} Can SIMQ support different graph-based ANNS backends?
\end{itemize}

\subsection{Experimental Setup}

\subsubsection{Datasets}

We evaluate our methods on five widely-used information retrieval datasets from the BEIR benchmark~\cite{thakur2021beir} and the ColBERTv2 evaluation suite~\cite{santhanam2022colbertv2}:

\begin{itemize}[leftmargin=*]
    \item \textbf{Quora}: A duplicate question retrieval dataset with 522,931 passages and 10K queries. Each query seeks semantically equivalent questions.

    \item \textbf{SciDocs}: A scientific paper retrieval dataset with 25,657 passages and 1K queries, where queries are paper titles and relevant passages are abstracts.

    \item \textbf{NQ (Natural Questions)}: A large-scale question answering dataset with 2,681,468 passages and 3.5K queries, derived from Google search queries and Wikipedia articles.

    \item \textbf{HotpotQA}: A multi-hop question answering dataset with 5,233,329 passages and 7.4K queries, requiring reasoning across multiple documents.

    \item \textbf{LoTTE-pooled}: A collection of five domain-specific forums (Writing, Lifestyle, Science, Technology, Recreation) with 2,819,103 passages and 3.9K queries, serving as an out-of-domain benchmark.
\end{itemize}

The statistics of datasets  are summarized in Table~\ref{tab:datasets}.

\begin{table}[t]
    \centering
    \caption{Dataset Statistics.}
    \label{tab:datasets}
    \small
    \setlength{\tabcolsep}{2pt}
    \begin{tabularx}{\columnwidth}{l|*{5}{>{\centering\arraybackslash}X}}
        \toprule
        \textbf{Dataset} & \textbf{Doc. tok.} & \textbf{Avg. tok./doc.} & \textbf{Queries} & \textbf{Avg. q. tok.} & $h$ \\
        \hline
        NQ             & 264.4M & 98.6  & 3.5K  & 11.9 & 64 \\
        LoTTE-pooled   & 389.5M & 138.2 & 3.9K  & 11.3 & 64 \\
        Quora          & 8.5M   & 16.3  & 10K   & 13.7 & 64 \\
        SciDocs        & 4.2M   & 164.4 & 1K    & 15.1 & 64 \\
        HotpotQA       & 349.1M & 66.7  & 7.4K  & 21.6 & 64 \\
        \bottomrule
    \end{tabularx}\vspace{-3ex}
\end{table}


\subsubsection{Baselines}

We compare SIMQ and SIMQ+ against the following state-of-the-art multi-vector retrieval methods:

\begin{itemize}[leftmargin=*]
    \item \textbf{PLAID}~\cite{santhanam2022plaid}: A multi-stage retrieval engine built on ColBERTv2, employing residual quantization and a centroid interaction matrix for candidate pruning.

    \item \textbf{MUVERA}~\cite{dhulipala2024muvera}: It reduces multi-vector retrieval to single-vector retrieval via Fixed Dimensional Encodings (FDEs) using LSH-based partitioning and random projections.

    \item \textbf{IGP}~\cite{bian2025igp}: A graph-based multi-vector retrieval method operating on the compressed centroid/residual representation. We use the PLAID-style compressed index and search the centroid graph before residual-decode reranking.

    \item \textbf{GEM}~\cite{tian2026gem}: A native graph-based index for multi-vector retrieval, which constructs proximity graphs over document vector sets using set-level clustering and bridges between local graphs.
\end{itemize}

All methods use the same ColBERTv2 token embeddings~\cite{santhanam2022colbertv2} with dimension $h = 64$ and the same evaluation qrels. Unless otherwise stated, we report Recall@100@100.

\subsubsection{Metrics}

We evaluate the following metrics:

\begin{itemize}[leftmargin=*]
    \item \textbf{Recall@$k$@$k$}: For each query, we divide the size of the intersection between the exact top-$k$ documents under MaxSim and the retrieved top-$k$ documents by $k$, then average this fraction over the evaluation queries. The main experiment reports $k\in\{20,50,100\}$; unless otherwise stated, the remaining experiments use $k=100$.

    \item \textbf{Queries Per Second (QPS)}: The number of queries the system can process per second, measuring throughput.


    \item \textbf{Index Construction Time}: The time required to build indexes, including any clustering or graph construction steps.

\end{itemize}

\subsubsection{Implementation and Hardware}

SIMQ and SIMQ+ are implemented in C++ with HNSW graph construction and search based on the hnswlib library.\footnote{\url{https://github.com/nmslib/hnswlib}} Except for PLAID's large-scale $k$-means centroid generation, the index-construction and query-processing experiments are conducted on a single-socket AMD EPYC 9T24 host with 96 physical cores, 192 logical threads, approximately 1.1~TiB memory, and local NVMe storage. SIMQ, SIMQ+, and IGP graph construction use 16 threads where applicable. The default is $\alpha=0.1$; the sensitivity sweep also evaluates $\{0.05, 0.08, 0.15, 0.20\}$.


\newcommand{\baselinepanel}[1]{\shortstack[c]{\includegraphics[width=0.19\textwidth]{#1}\\[0.3ex]{\scriptsize Recall}}}

\begin{figure*}[t]
	\centering
	\includegraphics[width=0.62\textwidth]{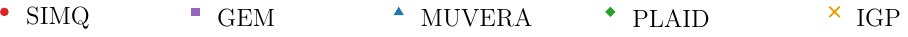}\\[-1ex]
	\subfloat[Quora, 20@20]{\baselinepanel{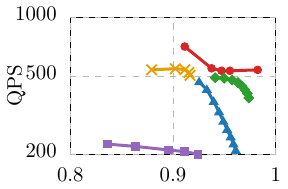}}\hfill
	\subfloat[SciDocs, 20@20]{\baselinepanel{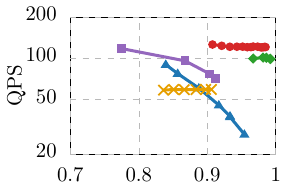}}\hfill
	\subfloat[LoTTE, 20@20]{\baselinepanel{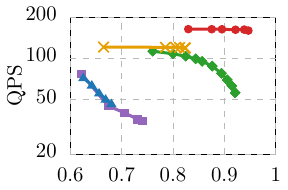}}\hfill
	\subfloat[NQ, 20@20]{\baselinepanel{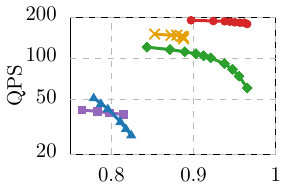}}\hfill
	\subfloat[HotpotQA, 20@20]{\baselinepanel{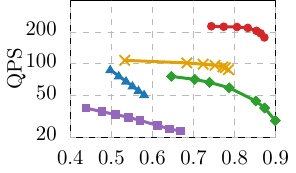}}\\[-2ex]
	\subfloat[Quora, 50@50]{\baselinepanel{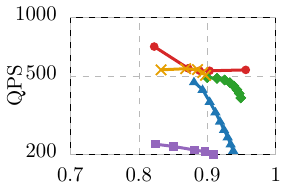}}\hfill
	\subfloat[SciDocs, 50@50]{\baselinepanel{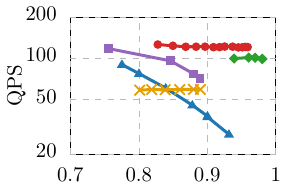}}\hfill
	\subfloat[LoTTE, 50@50]{\baselinepanel{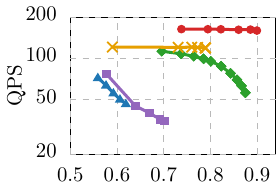}}\hfill
	\subfloat[NQ, 50@50]{\baselinepanel{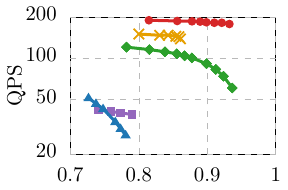}}\hfill
	\subfloat[HotpotQA, 50@50]{\baselinepanel{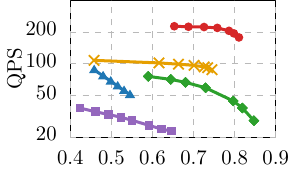}}\\[-2ex]
	\subfloat[Quora, 100@100]{\baselinepanel{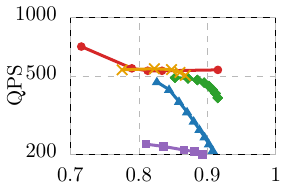}}\hfill
	\subfloat[SciDocs, 100@100]{\baselinepanel{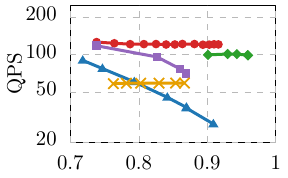}}\hfill
	\subfloat[LoTTE, 100@100]{\baselinepanel{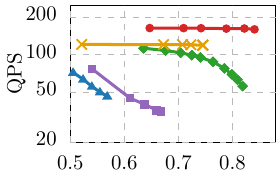}}\hfill
	\subfloat[NQ, 100@100]{\baselinepanel{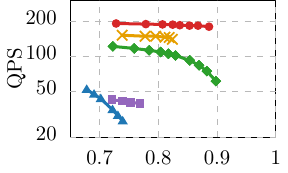}}\hfill
	\subfloat[HotpotQA, 100@100]{\baselinepanel{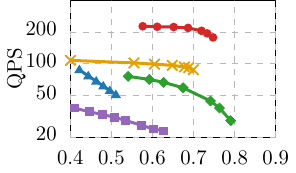}}
	\par\vspace{0ex}
	\caption{QPS versus Recall on five datasets.  QPS is on a logarithmic vertical axis.}\vspace{-2ex}
	\label{fig:qps_recall}
\end{figure*}

\subsubsection{Parameter Settings}

\begin{itemize}[leftmargin=*]
    \item \textbf{PLAID}: 4-bit residual quantization (threshold 0.3) with $K = 7.5\text{k}$ (SciDocs), $15\text{k}$ (Quora), and $262.1\text{k}$ (others). Experiments sweep the number of centroid cells probed per query token $n_{\mathrm{cells}}$ from 1 to 25 at the candidate-document filtering budget $n_{\mathrm{docs}} = 4\text{k}$, retaining 1,000 documents for exact MaxSim reranking.

    \item \textbf{MUVERA}: The FDE uses $R=20$ independent partition repetitions, $b=4$ SimHash bits per partition, and projected dimension $d_{\mathrm{proj}}=16$ per cell. The DiskANN index~\cite{subramanya2019diskann} uses maximum graph degree $R_d=32$, construction-search candidate-list size $L=100$, and edge-pruning factor $\alpha_e=1.2$. Both constructions use 16 threads.

    \item \textbf{SIMQ}: Sampling ratio $\alpha = 0.1$, maximum cluster size $T = 160$, and two children per recursive split ($K=2$, local to this splitting step). The multi-threaded construction uses deferred recursive splitting after token assignment. The HNSW graph uses degree parameter $M=32$ and construction-search candidate-list size $\mathit{efConstruction}=50$. The per-token candidate-document budget is $n_{\mathrm{d}}=1000$; $n_c$ separately counts the centroids retrieved per query token. The exact-reranking budget is $k_{\mathrm{r}}=1000$.

    \item \textbf{SIMQ+}: We use the same index, graph-search, and execution settings as SIMQ, with minimum candidate size $N_{\min}=1000$. 

    \item \textbf{GEM}: The number of clusters probed during search is $n_{\mathrm{probe}}=1$--$2$. Its dataset-specific $K$ and 20-iteration $k$-means settings follow PLAID.

    \item \textbf{IGP}: PLAID-style 4-bit residual-compressed index, centroid HNSW graph built with 16 threads, and residual-decode reranking. The parameter $\mathrm{probe\_topk}=1400$ controls the candidate-document search budget. The parameter $n_{\mathrm{probe}}$ controls the number of clusters probed; the Quora sweep uses $n_{\mathrm{probe}}=1$--$20$.
\end{itemize}


\subsection{Experimental Results}

\subsubsection{RQ1: Comparison with Baselines}

\paragraph{QPS-Recall Performance.}
Figure~\ref{fig:qps_recall} shows SIMQ and baseline QPS--Recall curves across five datasets for 20@20, 50@50, and 100@100 retrieval settings.

SIMQ and PLAID have different query-time trade-offs. PLAID computes a full $K\times m$ centroid--query interaction matrix, where $K$ is the total centroid count in its residual-compressed representation. Larger $K$ yields finer clusters but increases matrix computation; smaller $K$ reduces this cost but coarsens clusters and can increase candidate-processing work. At fixed $K$, probing fewer cells does not remove the full matrix cost, while probing more cells increases downstream document-scoring work. SIMQ instead navigates directly to relevant centroids through per-token graph search, supporting fine-grained partitions without a full centroid scan, thus can beat the baselines in query performance.

MUVERA's LSH cell quality depends on how its random hyperplanes align with non-uniform token structures; while $R=20$ repetitions reduce variance, the partition remains data-oblivious. In contrast, SIMQ's PAG-based partitioning adapts directly to the token distribution.


GEM~\cite{tian2026gem} builds proximity graphs over document vector sets via a costly pipeline (set-level clustering, local graphs, and bridging). IGP adds a centroid graph to PLAID's representation, reporting build overhead relative to PLAID. All methods are evaluated using Pareto frontiers rather than single operating points.


\paragraph{Sampling Ratio Sensitivity.}
To assess the effect of the sampling ratio, we compare SIMQ with $\alpha$ values 0.05, 0.08, 0.1, 0.15, and 0.20 on NQ and Quora. The default setting is $\alpha=0.1$. For each setting, we vary the retrieval parameter and plot the measured QPS--Recall@100@100 operating points.

\begin{table}[t]
	\centering
	\caption{Total Index Construction Time (seconds). The fastest reported method for each dataset is shown in bold.}
	\label{tab:build_time}
	\small
	\begin{tabular}{l|rrrr}
		\toprule
		\textbf{Dataset} & \textbf{PLAID} & \textbf{IGP} & \textbf{MUVERA} & \textbf{SIMQ} \\
		\hline
		NQ             & / & / & $\mathbf{2{,}458}$ & 3,386 \\
		LoTTE-pooled   & / & / & $\mathbf{2{,}936}$ & 5,271 \\
		Quora          & 4,282 & 4,282.3 & 264 & $\mathbf{68}$ \\
		SciDocs        & 1,061 & 1,061.2 & $\mathbf{21}$ & 33 \\
		HotpotQA       & / & / & 4,058 & $\mathbf{4{,}020}$ \\
		\bottomrule
	\end{tabular}
\end{table}

\begin{figure}[t]
    \centering
    \includegraphics[width=0.82\columnwidth]{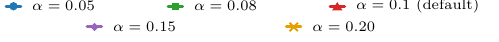}\\[-3ex]
    \subfloat[NQ across sampling ratios]{\makebox[0.495\columnwidth][c]{\includegraphics[height=0.95in]{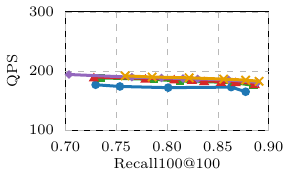}}}\hfill
    \subfloat[Quora across sampling ratios]{\makebox[0.495\columnwidth][c]{\includegraphics[height=0.95in]{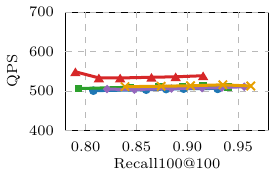}}}
    \caption{QPS--Recall@100@100 curves for SIMQ.}\vspace{-1ex}
    \label{fig:simq_alpha_sensitivity}
\end{figure}

Figure~\ref{fig:simq_alpha_sensitivity} shows that SIMQ maintains broadly comparable QPS--Recall trade-offs for sampling ratios $\alpha$ between 0.08 and 0.20 on both datasets, with the default $\alpha=0.1$ in this stable range. The smallest setting, $\alpha=0.05$, is less robust; its throughput lags at the high-recall end on NQ and offers no consistent advantage on Quora. This trend is consistent with the effect of sparse sampling: fewer sampled centroids lead to larger clusters, whose greater radii can make approximate scores less discriminative for identifying target documents. Overall, the results support $\alpha=0.1$ as a practical default and suggest avoiding overly small sampling ratios.

\paragraph{Index Construction Time.}
Table~\ref{tab:build_time} reports the total index construction time for each dataset and method. PLAID uses 20 $k$-means iterations. SciDocs and Quora were small enough to complete this construction on the CPU host; for the larger datasets, the $k$-means construction was completed on a separate GPU machine and no CPU construction time is reported. GEM is omitted because its construction cost is not comparable under the present CPU setup.

The slash indicates that no CPU construction time is reported; the corresponding PLAID indexes were completed using $k$-means on a separate GPU machine, so IGP total time is also unavailable. IGP times include the corresponding PLAID construction and centroid-graph construction overhead. MUVERA times include FDE encoding, DiskANN construction, and index serialization. GEM is not included in this table because its construction is substantially more expensive under the CPU configuration. The fastest reported construction time in each row is bolded.

\subsubsection{RQ2: PAG vs.\ Alternative Clustering Methods}

To compare the partitioning strategies, we evaluate PAG, $k$-means, and LSH on all five datasets using the same downstream graph and query pipeline. $k$-means is a useful reference because, with sufficiently many centroids, it directly optimizes a data-aware clustering objective. We test whether PAG's sampling- and graph-based construction can preserve retrieval behavior while reducing the CPU burden of repeated global $k$-means iterations. For PAG and $k$-means, we select measured operating points with nearly matched Recall@100@100; no interpolated values are used. LSH is reported at its available measured operating point because its recall range is lower and some datasets have only a single completed sweep point. Its occupied-cell count and cluster sizes are data-dependent, which can lead to uneven partition utilization.

\begin{table}[t]
    \centering
    \caption{PAG and $k$-means at nearly matched Recall@100@100, with LSH shown at its available operating point. Each cell reports QPS / Recall@100@100.}
    \label{tab:clustering_compare}
    \small
    \setlength{\tabcolsep}{3pt}
    \begin{tabular}{l|ccc}
        \hline
        \textbf{Dataset} & \textbf{PAG} & \textbf{$k$-means} & \textbf{LSH} \\
        \hline
        SciDocs & 121.6 / 0.787 & 105.0 / 0.794 & 106.4 / 0.480 \\
        Quora & 532.7 / 0.834 & 516.2 / 0.840 & 504.0 / 0.421 \\
        LoTTE-pooled & 163.4 / 0.710 & 160.6 / 0.684 & 148.1 / 0.314 \\
        NQ & 187.6 / 0.779 & 183.8 / 0.771 & 169.2 / 0.371 \\
        HotpotQA & 227.1 / 0.576 & 216.8 / 0.565 & 214.6 / 0.262 \\
        \hline
    \end{tabular}
\end{table}

Across the five datasets, PAG achieves recall close to $k$-means and comparable or higher QPS at the selected operating points. Its construction is more CPU-friendly by design: PAG samples aggregation points and assigns residual vectors through graph-guided search instead of repeatedly applying global Lloyd iterations to all token vectors; only oversized clusters undergo local recursive two-way splitting. Table~\ref{tab:clustering_compare} reports retrieval throughput and recall, not a direct measurement of clustering construction time. LSH is included only at its available measured point, where its recall is substantially lower, so it is not a recall-matched comparison.

\subsubsection{RQ3: SIMQ vs.\ SIMQ+}
We vary $n_c$, the number of nearest centroids retrieved per query token, to compare SIMQ and SIMQ+ on LoTTE, NQ, HotpotQA, and Quora. SIMQ+ counts all documents in the retrieved clusters before document selection, rather than applying basic SIMQ's per-token approximate-score truncation. It uses hit-count ranking and expands tied candidates at the dynamic boundary before exact reranking.

\begin{figure}[t]
    \centering
    \includegraphics[width=0.48\columnwidth]{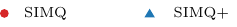}\\[-2ex]
    \subfloat[LoTTE]{\includegraphics[width=1.65in]{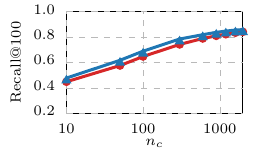}}\hfill
    \subfloat[NQ]{\includegraphics[width=1.65in]{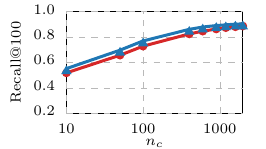}}\\[-2ex]
    \subfloat[HotpotQA]{\includegraphics[width=1.65in]{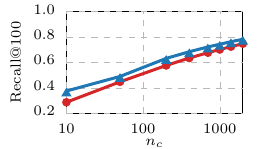}}\hfill
    \subfloat[Quora]{\includegraphics[width=1.65in]{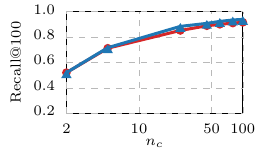}}
    \caption{SIMQ vs.\ SIMQ+ Recall@100@100 as a function of $n_c$.}\vspace{-1ex}
    \label{fig:simq_simqplus}
\end{figure}

SIMQ+'s gains stem from two design changes: a) Count-based scoring: replaces weak centroid inner-product scoring with distinct query-token hit counts across the retrieved pool, avoiding SIMQ's per-token score truncation; b) Dynamic candidate selection: gathers candidates by descending hit count until reaching at least $N_{\min}$ documents, retaining all tied candidates at the boundary to prevent arbitrary exclusion.


\subsubsection{RQ4: Flexibility across Graph Backends}

SIMQ treats the centroid graph as a replaceable ANNS backend: any graph-based backend that indexes the cluster centroids and supports nearest-centroid search can provide the navigation stage of the framework. HNSW is the default implementation used in the main experiments, rather than a requirement of SIMQ. To demonstrate this flexibility, we evaluate three representative backends: HNSW~\cite{malkov2020hnsw}, NSG~\cite{fu2019fast}, and DPG~\cite{li2020dpg}, on NQ and HotpotQA. We retain the same PAG-based partitioning, candidate generation, and exact MaxSim reranking pipeline, replacing only the graph construction and search algorithms. All backends use deferred recursive $K=2$ partition refinement after token assignment.

We keep the shared retrieval parameters fixed across backends, so the comparison examines substituting the centroid-navigation backend within SIMQ.

\begin{figure}[t]
    \centering
    \includegraphics[width=0.82\columnwidth]{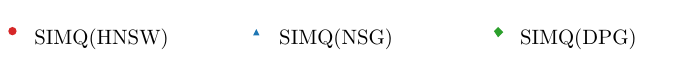}\\[-2ex]
    \subfloat[NQ]{\includegraphics[height=0.78in]{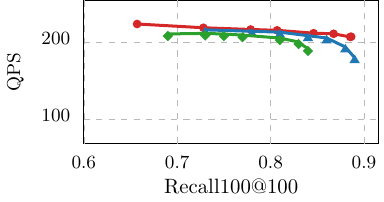}}\hspace{1.2em}
    \subfloat[HotpotQA]{\includegraphics[height=0.78in]{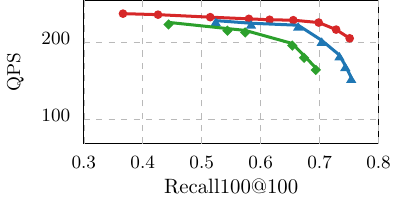}}
    \caption{QPS--Recall@100@100 comparison across graph backends.}\vspace{-1ex}
    \label{fig:index_single_multi}
\end{figure}

Figure~\ref{fig:index_single_multi} shows that all three backends support SIMQ's retrieval pipeline, demonstrating that its centroid navigation is not tied to HNSW. Their curves exhibit different recall--throughput trade-offs, so backend substitution preserves the framework while allowing different performance characteristics. HNSW, NSG, and DPG provide representative empirical validations of this modular design; the reported performance comparisons concern the measured NQ and HotpotQA results.

\subsection{Summary of Experimental Findings}

The experimental results lead to the following conclusions:

\begin{enumerate}
    \item SIMQ and SIMQ+ provide competitive retrieval quality and efficiency against PLAID, MUVERA, IGP, and GEM across five diverse datasets; exact operating-point claims are reported in the corresponding figures and tables.

    \item At nearly matched Recall@100@100, PAG achieves retrieval quality close to $k$-means with comparable or higher QPS. Its sampling and graph-guided assignment provide a more CPU-friendly construction path than repeated global $k$-means iterations; Table~\ref{tab:clustering_compare} reports retrieval metrics, while LSH is shown only at its available lower-recall operating point.

    \item SIMQ+'s count-based scoring and dynamic candidate set address ranking reversals caused by biased approximate scores under fine-grained partitioning.

    \item SIMQ separates its multi-vector retrieval pipeline from the underlying centroid-navigation backend. Experiments with HNSW, NSG, and DPG on NQ and HotpotQA demonstrate that the backend can be replaced without changing partitioning, candidate generation, or exact reranking; the recall--throughput trade-off varies with the chosen backend.
\end{enumerate}

\section{Related Work}
\label{sec:relatedWork}

\stitle{Multi-Vector Retrieval Models.} ColBERT~\cite{khattab2020colbert} and ColBERTv2~\cite{santhanam2022colbertv2} use token-level MaxSim scoring, which is computationally expensive. PLAID~\cite{santhanam2022plaid} accelerates this via $k$-means centroids and residual quantization (also adopted by XTR~\cite{lee2023xtr} and WARP~\cite{scheerer2025warp}), though it faces bottlenecks in matrix granularity and GPU-dependent clustering. Alternatively, MUVERA~\cite{dhulipala2024muvera} compresses documents into fixed-dimensional encodings (FDEs) via LSH and random projections; however, its data-oblivious partitioning requires $R$ hash repetitions to stabilize recall.


\stitle{Graph-based Multi-vector Retrieval.} GEM~\cite{tian2026gem} builds a dual proximity graph over document vector sets, decoupling Earth Mover's Distance based construction from Chamfer scoring via semantic shortcuts, though its set-to-set computations scale with token length. IGP~\cite{bian2025igp} builds a proximity graph over PLAID centroids using incremental probing for candidate generation while retaining PLAID's quantization. VecFlow-Chamfer~\cite{vecflow2026} is a GPU-optimized multi-vector retrieval method for high-throughput systems.


\stitle{Data Partitioning for Retrieval.} Partitioning approaches vary: PLAID uses data-aware yet GPU-heavy $k$-means clustering, while MUVERA relies on fast but data-oblivious LSH. DSANN~\cite{yu2025dsann} introduces the Point Aggregation Graph (PAG) via sampling for data-aware partitions without iterative clustering. SIMQ adapts PAG for multi-vector retrieval, delivering fine-grained, stable clusters without GPU dependency.


\stitle{Graph-based ANNS Methods.}
Graph-based indices such as HNSW~\cite{malkov2020hnsw}, Vamana~\cite{subramanya2019diskann},
and NSG~\cite{fu2019fast} are widely used graph-based methods for single-vector approximate nearest
neighbor search. SIMQ builds
upon these single-vector graph methods by constructing a graph index over cluster centroids
and adapting per-token graph navigation to approximate multi-vector MaxSim scoring.

\section{Conclusion}
\label{sec:conclusion}

We presented SIMQ, a multi-vector retrieval framework that combines data-aware
sampling-based clustering with per-token graph navigation. By indexing cluster
centroids refined after sampling-based initialization, SIMQ avoids materializing
a centroid interaction matrix and reduces reliance
on GPU-based large-scale $k$-means. Under a manifold assumption and successful cell
retrieval, our analysis shows that SIMQ has lower expected intra-cell approximation
error than MUVERA's data-oblivious LSH partitioning. SIMQ+ adds count-based scoring
and a dynamic candidate set to improve ranking robustness. Experiments on five
benchmarks measure retrieval quality, throughput, and construction-time trade-offs
against PLAID, MUVERA, IGP, and GEM, and evaluate alternative graph backends on CPUs.

%


\newpage

\section*{Acknowledgment}
During the preparation of this work, the authors used OpenAI Codex for coding
assistance. The authors
are responsible for reviewing and validating all AI-assisted material and take
full responsibility for the final manuscript, implementation, and reported results.

\balance
\bibliographystyle{ieeetr}
\bibliography{add}
\end{document}